\magnification=\magstep1
\hfuzz=6pt
\baselineskip=15pt
$ $
\vskip 1in
\centerline{\bf Universe as quantum computer}
\bigskip
\centerline{Seth Lloyd}

\centerline{d'Arbeloff Laboratory for Information Systems
and Technology}

\centerline{MIT 3-160, Cambridge, Mass. 02139}

\centerline{{\it and} The Santa Fe Institute}

\centerline{slloyd@mit.edu}

\bigskip
\noindent{\it Abstract:} This paper shows that
universal quantum computers possess decoherent histories
in which complex adaptive systems evolve with high probability.

\vskip 1cm
\noindent{\bf Introduction} 

The universe is quantum mechanical.$^1$
It begins in some quantum state,
evolves by dynamics specified by a quantum
time-evolution operator, 
and by the process of decoherence
generates the quasi-classical world that we see around 
us,$^{1-4}$
a world filled with structure and complexity. 
This paper addresses the question of whether this observed
structure and complexity arose naturally, or whether they
constitute an extremely improbable accident.  Rather than treating
the actual universe in all its complexity (which would 
be hard), this paper investigates a `toy' model of a
quantum universe --- a quantum computer --- and shows 
that even simple quantum dynamical systems evolve complexity
and structure naturally and with high probability.

\bigskip\noindent{\bf 1. Decoherence} 

Decoherence is a
process whereby the underlying dynamics of a system select
out certain quantum variables, e.g. hydrodynamic variables,
and prevent them from exhibiting coherent behavior such as
interference.  Such decoherent variables can be described by
the ordinary classical laws of probability to a high degree of
precision.$^{1-4}$  Decoherence plays a key role in shaping the
quantum universe: in addition to determining 
quasi-classical behavior, it elevates some quantum
fluctuations to the status of `frozen accidents' that 
substantially effect the subsequent course of events.  
The features of our universe that result from frozen
accidents are exactly those that are not pre-programmed by
the basic dynamics of the universe and the initial state.
An example of such a frozen accident is the fluctuation
in the primordial density of matter that put our
galaxy here rather than elsewhere.  Features of our
universe that could have resulted from frozen accidents
include living creatures' use of right-handed sugars rather
than left-handed sugars, the masses and coupling constants
in the equations of motion for the elementary particles,
and the particular mix of your mother's and father's genes
that make you you rather than your sister.  
To create a frozen accident that has a significant effect on
your life, use a quantum event such as a radioctive decay to
generate a random bit of information and use that bit to
decide whether to buy a Ford or a Chevrolet.

\bigskip\noindent{\bf 2. Quantum Computers} 

To understand
how decoherence can lead to the spontaneous generation of
complexity, consider a quantum computer.  A quantum computer
is a quantum system that computes.$^{5-6}$ 
A large variety of quantum systems are
capable of quantum computation in principle.$^7$
A simple example of a quantum computer
is a quantum system with states $|b\rangle$ corresponding
to bit strings $b$ with a unitary time evolution $U$ that
gives
$$U|b\rangle=|\Upsilon(b)\rangle\quad,\eqno(1)$$
\noindent where $\Upsilon(b)$ is the image of the logical
state $b$ under the action of a universal {\it classical}
computer.$^8$  To conform with the usual laws of quantum
mechanics, $\Upsilon$ must be one-to-one; 
to conform with the usual laws of computation, $\Upsilon$
involves only local logical operations that act on a few
bits at a time.  (This quantum computer is simply a
reversible classical computer recast in a quantum picture;
to get a general quantum computer that can perform
non-classical operations such as creation of superpositions
and entanglement, one must adjoin to the ordinary classical
logic operations such as {\it AND}, {\it OR} and {\it NOT}
some simple quantum logic operation such as the rotation of
a quantum bit.$^9$  Such more general quantum
computers will be considered below.)  

Because its underlying logical operation is deterministic,
the computations carried out by the quantum computer of 
equation (1) 
correspond to decoherent histories.  The conditions under
which a particular set of quantum variables are decoherent
and can be assigned probabilities were given in references
(1,3-4).  To verify that the sequence of logical
states followed by the quantum computer decohere, note that
$$\eqalign{D=&~{\rm tr} P(b_{n+1})U P(b_n)\ldots U P(b_1) 
|\psi\rangle\langle\psi|
P(b'_1)U^\dagger\ldots  P(b'_n)U^\dagger\cr
 \propto&~
\delta_{b_1b'_1} 
\delta_{b_2\Upsilon(b_1)} \ldots
\delta_{b_nb'_n}
\delta_{b_{n+1}\Upsilon(b_n)}
,\cr}\eqno(2)$$
\noindent where the $P(b)=|b\rangle\langle b|$ are
projection operators on the logical states of the
computer and  $|\psi\rangle = \sum_b \psi_b|b\rangle$
is an arbitrary initial state.
$D$ is called the decoherence functional: the histories
composed of sequences of logical states $b_1\ldots b_n$ can
be assigned probabilities that obey the usual classical sum
rules for probabilities if and only if the off-diagonal
terms of $D$ are purely imaginary.  Here the deterministic
nature of $\Upsilon$ implies that these off-diagonal
terms vanish, so the histories decohere.  The on-diagonal
terms of $D$ give the probabilities for the different
histories: these terms vanish unless $b_{n+1}=\Upsilon(b_n)
= \Upsilon^2(b_{n-1})= \ldots = \Upsilon^n(b_1)$, assigning
non-zero probability only to legitimate computations.

Equation (2) implies that the computations performed
by a quantum computer decohere, but it gives
an unrealistic picture of the
introduction of frozen accidents: all quantum fluctuations
are frozen in at the first time step, and the action of the
computer is completely deterministic thereafter.  To give a
more realistic picture of how frozen accidents occur, recall
that decoherence is a dynamic process in which interactions
between different parts of a quantum system render certain
variables effectively classical.  Let 
$P_{\rm loc}(b)$ be projection operators onto the bits  
of $b$ that were involved in the local action 
of $\Upsilon$ at the previous time step: 
i.e., $P_{\rm loc}(b)$ acts only on the bits ${\rm loc}(b)$ of $b$
that have undergone a non-trivial dynamics --- they act only
where there is interaction.  (Here `local' means
that the way in which the computer updates a given bit 
of information depends only on a {\it finite} number 
of bits in its memory.  This local quality of information 
processing stems from the local nature of the underlying
physical interactions that determine the dynamics of
the computer.)
Now look at the {\it local}
decoherence functional
$$\eqalign{
D_{\rm loc}=&~{\rm tr} P_{\rm loc}(b_{n+1})U 
P_{\rm loc}(b_n) \ldots U P_{\rm loc}(b_1) 
|\psi\rangle\langle\psi|
P_{\rm loc}(b'_1)U^\dagger\ldots P_{\rm loc}(b'_n))
U^\dagger\cr
 \propto&~
\delta_{{\rm loc}(b_1){\rm loc}(b'_1)} 
\delta_{{\rm loc}(b_2) {\rm loc}(\Upsilon(b_1))}\ldots
\delta_{{\rm loc}(b_n){\rm loc(b'_n)}} 
\delta_{{\rm loc}(b_{n+1}) {\rm loc}(\Upsilon(b_n)},
\cr}\eqno(3)$$
\noindent Equation (3) describes the decoherence process
for the {\rm local} histories of the bits in the computer
that have undergone a non-trivial dynamics. 
Since the off-diagonal terms of $D_{\rm loc}$
vanish, these local histories of the quantum computer also
decohere, and have non-zero probability if and only if they
obey the proper local dynamics given by $\Upsilon$. 

The local histories of equation (3) give a more realistic
picure of how decoherence occurs: bits that 
undergo logical transformations decohere via their local
interactions.  Frozen accidents are introduced locally
at a variety of times, and where and when they occur 
depends on the previous history of the computation, so that
the results of frozen accidents in the past determine the
complexion of frozen accidents in the future. 
Equation (3) distinguishes only between computations
that have processed different information in the past,
and not between computations that differ only in some as 
yet to be read bits in the future.  Bits that have not
yet been read constitute frozen accidents that are 
waiting to happen.

(When the quantum computer allows the systematic
creation and manipulation of superpositions the
computational histories do not
automatically decohere.  In Shor's algorithm for factoring,$^{10}$
for example, the computer creates a superposition of
logical inputs and then performs the same mathematical 
manipulations on each term in the superposition.  The
histories got by projecting onto the individual terms of
this superposition together with the final answer to the
factoring problem are not decoherent.  The decoherence
functional formalism can still be applied to such computers,
but care must be taken to include in $D$ only
projections onto subsets of bits whose values are
correlated with the value of some subset of bits 
at the time of the projection onto the final
result of the computation.  In the formalism
of Gell-Mann and Hartle,$^1$   
the value of the future bits then form
a `record' of the value of the bits in the past
and guarantee decoherence.  The discussion above then
holds for these more general quantum computers as well.)

\bigskip\noindent{\bf 3. Quantum computers and the
generation of complexity} 

Sections 1 and 2 investigated the ideas of decoherence as
applied to a simple model of quantum computation.
Quantum computers possess a natural set of decoherent
histories --- the local histories defined above --- that
decohere through local interactions and that exhibit the
phenomenon of frozen accidents.  The question that this
paper set out to answer is, To what extent do these
histories describe the evolution of complex structures?

Remarkably, the answer to this question is, Not only do
quantum computers evolve complex structures naturally and
with high probability, they evolve {\it all conceivable}
structures, simple and complex, starting from an arbitrary
initial state $|\psi\rangle$.  An arbitrary initial state
represents all finite bit strings $b$ with approximately equal 
probability.  The total probability over all bit strings
that commence with a particular $\ell$-bit program 
is then $2^{-\ell}$.  That is, the
probabilities for different computational histories 
correspond to a computer that
has been programmed with a random program.  

At first it might seem that a computer programmed at random
is likely to generate gibberish: garbage in, garbage out.
After all, monkeys typing random texts on typewriters have
a finite probability of producing the works of Shakespeare;
it's just that the probability of producing any given long bit
string of length $N$ from a random process is $2^{-N}$ and
goes exponentially to zero as $N\rightarrow\infty$.
Since the works of Shakespeare take more than a million
bits to write down in binary form, the monkeys have
a chance of less than $2^{-1,000,000}\approx 1/10^{300,000}$
of getting them right, which is not worth losing sleep over.  
The point is that a long random bit string is highly unlikely
to exhibit complex or interesting structure.  As the
length of the description of some desired structure goes
to infinity, the probability of producing it goes
exponentially to zero.

In contrast, a computer 
programmed at random has a finite probability for 
producing any desired complex or interesting structure.
As Bennett notes,$^{11}$ a computer programmed at random
can be thought of as the result of monkeys typing
random texts into a BASIC interpreter rather than
monkeys typing on a typewriter.
For a computer, there is a big difference between a 
random input and a random output.
Take for instance the first million bits of the binary
expression for $\pi$.  This is a sufficiently interesting
number that people have devoted considerable resources
to calculating it.  The chances of this
number appearing as the first million bits of a 
random bit string are the about the same as for
the monkeys typing out the works of Shakespeare.
The chances of a randomly programmed computer producing
the first million bits of $\pi$, by comparison, are
at least as great as $2^{-K(\pi)}$, where $K(\pi)$ is the
length of the shortest computer program that generates
$\pi$ ($K(\pi)$ is called the 
algorithmic information content$^{12}$ of $\pi$).  
For a conventional computer $K(\pi)$ is
likely to be no more than a few hundred bits, leading
to a probability of producing the first million
bits of $\pi$ approximately $10^{3000}$ higher by typing into
a computer than by typing into a typewriter.
For a special-purpose
computer with built in modules for simulating
hydrodynamics or general relativity, the probability
of computing $\pi$ is even greater: almost any
initial conditions will lead to structures such
as spherical droplets or elliptical orbits that
effectively embody $\pi$ to high precision.

The contrast between random inputs and random outputs
can be made sharper by looking at the probability
of producing structures described by arbitrarily long bit
strings.  Let $s$ be an infinite bit string describing
the infinite extension of some complex structure. 
No matter how many bits of $s$ are
required, the randomly programmed computer has
the same finite probability, $2^{-K(s)}$, of producing
them, while the probability of producing the bits
by a random process such as coin-tossing
goes rapidly and exponentially to zero.

Clearly, a quantum computer that starts in a random
state has a finite probability of producing not only $\pi$,
but any other algorithmically specifiable bit string
as well.  The computer favors bit strings that have
low algorithmic information content, i.e., strings
that are highly non-random.  Complex bit strings
form a finite-probability subset of the set of
non-random strings.  Here what is meant by `complex'
is largely up to the reader.   For example, one could
take complex strings to be those
that are logically deep in the sense of 
Bennett$^{11}$, or those that are thermodynamically 
deep in the sense of Lloyd and Pagels$^{13}$, or
those that are effectively complex in the sense
of Gell-Mann$^{14}$.  As long as a complex structure,
even one of infinite size, can be specified by a 
finite computer program then it has a non-zero
probability of being produced by a quantum computer
starting from a random state.
The relative probability of different outputs
for the computer depends on the computer's dynamics,
but differs from computer to computer by at most a
constant factor.  Quantum computers exhibit naturally
decohering histories that tend to evolve complex
structures.

\bigskip\noindent{\bf 4. The universe in the computer}

What do the decoherent histories generated by a universal 
quantum computer contain?  They contain 
computations that generate all algorithmically 
specifiable structures, including digital
complex adaptive systems.  If atoms,
molecules, and biological structures
can be simulated by a quantum computer
(and there is no reason to think that cannot),
that computer contains digital analogues of
you and me.  

Is the universe that we see around us
of the sort described above?  Perhaps, but only with
some subtleties taken into account.
The dynamics of our universe are given by a local
Hamiltonian specified by quantum field theory,
not by the local unitary dynamics of a quantum computer.
A system with a local unitary dynamics
cannot have a local Hamiltonian dynamics.$^{15}$  As Feynman
noted, however,$^{16}$ a quantum computer with the unitary
dynamics given by $U$ above has the same computational
histories as a quantum computer with the Hamiltonian
dynamics given by $H=U+U^\dagger$.  It is straightforward
to verify that these histories are decoherent. 
Our universe could indeed be some type of highly parallel
Hamiltonian quantum computer: whether or not it is could in
principle be verified by constructing a lattice version
of the standard model, and checking whether the resulting
dynamics supports computation.  If so, then the results of
this paper imply that it is not
surprising that the universe is so complex.

\vfill
\noindent{\it Acknowledgements:} This work was supported 
by ONR and by DARPA/ARO under the intiative for 
Quantum Information and Computation (QUIC).
The author would like to thank Sidney Coleman, Murray
Gell-Mann, and Jim Hartle for helpful comments.
\eject

\noindent{\bf References:}

\item{1.} M. Gell-Mann, J. Hartle, in {\it Complexity,
Entropy, and the Physics of Information}, W.H. Zurek ed.,
Santa Fe Institute Studies in the Sciences of Complexity
{\bf VIII}, Addison-Wesley, Redwood City, 1990.  
{\it Phys. Rev. D} {\bf 47}, 3345 (1993); in {\it Proceedings of the
4th Drexell symposium on quantum non-integrability --- The
quantum-classical correspondence}, D.-H. Feng, ed., in press.  

\item{2.} W.H. Zurek {\it Physics Today} {\bf 44},
36, (1991); {\it Phys. Rev. D} {\bf 24}, 1516 (1981);
{\it Phys. Rev. D} {\bf 26}, 1516 (1981).

\item{3.} R. Griffiths, {\it J. Stat. Phys.} {\bf 36},
219 (1984).

\item{4.} R. Omnes, {\it J. Stat. Phys.} {\bf 53}, 893
(1988).  {\it Ibid.}, 933.  {\it Ibid.}, 957.

\item{5.} D. Divincenzo, {\it Science} {\bf 270}, 255 (1995).

\item{6.} S. Lloyd, {\it Sci. Am.} {\bf 273}, 140 (1995).

\item{7.} S. Lloyd, {\it Phys. Rev. Lett.} {\bf 71}, 943 (1993);
{\it J. Mod. Opt.} {\bf 41}, 2503 (1994).

\item{8.} P. Benioff, {\it J. Stat. Phys.} {\bf 22}, 563 (1980);  {\it
Phys. Rev. Lett.} {\bf 48}, 1581 (1982);  {\it J. Stat. Phys.} {\bf
29}, 515 (1982);  {\it Ann. N.Y. Acad. Sci.} {\bf 480}, 475 (1986).

\item{9.} D. Deutsch, {\it Proc. R. Soc. London Ser. A} {\bf 400},
97 (1985).  

\item{10.} P. Shor, in {\it Proceedings of the 35th Annual Symposium on
Foundations of Computer Science}, S. Goldwasser, ed., ({\it IEEE} Computer
Society, Los Alamitos, CA, 1994), p. 124.

\item{11.} C.H. Bennett, in {\it Emerging Syntheses in Science},
D. Pines, ed., Santa Fe Institute Volume I, Addison Wesley,
Redwood City (1988).

\item{12.} R.J. Solomonoff, {\it Inf. \& Contr.} {\bf 7}, 1 (1964). 
A.N. Kolmogorov, {\it Inf. Trans.} {\bf 1}, 3 (1965).
G.J. Chaitin, {\it J. ACM} {\bf 13}, 547 (1966).

\item{13.} S. Lloyd and H. Pagels, {\it Ann. Phys.} {\bf 188},
186-213, 1988.

\item{14.} M. Gell-Mann, {\it The Quark and the Jaguar},
W.H. Freeman, New York, 1994; {\it Complexity} {\bf 1}, 16 (1995).
M. Gell-Mann and S. Lloyd, {\it Complexity} {\bf 2/1}, 44 (1996).

\item{15.} S. Coleman, private communication.  When the unitary
time evolution $U$ over a discrete finite time $\Delta t$ is 
local, then the time evolution over a fractional time, 
e.g.,  $\Delta t/2$, is non-local, as can be seen by 
decomposing $U$ in terms of eigenstates as in reference 7.  

\item{16.} R.P. Feynman, {\it Opt. News} {\bf 11}, 11 (1985); {\it
Found. Phys.} {\bf 16}, 507 (1986).  See also
N. Margolus, {\it Ann. N.Y. Acad. Sci.} {\bf 480}, 487
(1986); and in {\it Complexity, Entropy, and the Physics of
Information}, W.H. Zurek, ed. (Santa Fe Institute Series, vol. 8,
Addison Wesley, Redwood City CA 1991), pp. 273--288.

\vfill\eject\end